\documentclass[prb,sd,amsmath,amssymb,reprint]{revtex4-1}

\usepackage{graphicx}
\usepackage{bm}
\usepackage{color}
\usepackage{here}

\begin{document}


\title{Comprehensive search for buckled honeycomb binary compounds based on noble metals (Cu, Ag, and Au)}

\author{Shota Ono}
\email{shota\_o@gifu-u.ac.jp}
\affiliation{Department of Electrical, Electronic and Computer Engineering, Gifu University, Gifu 501-1193, Japan}

\begin{abstract}
Honeycomb structure has been frequently observed in two-dimensional (2D) materials. CuAu in the buckled honeycomb (BHC) structure has been synthesized recently, which is the first case of 2D intermetallic compounds. Here, the dynamical stability of 2D $AX$ in the BHC structure, where $A=$ Cu, Ag, and Au and $X$ is a metallic element in the periodic table, is systematically studied by calculating phonon dispersions from first-principles. Among 135 $AX$, more than 50 $AX$ are identified to be dynamically stable. In addition, (i) a relationship between the dynamical stability and the formation energy, (ii) a correlation of dynamical stability between different constituents $A$, (iii) a trend of lattice parameters, and (iv) electronic and magnetic properties are discussed. Furthermore, a stable phase of B11-type AuZr is predicted based on both the result (ii) and the stability relationship between 2D and three-dimensional structures. The present findings stimulate future studies exploring physics and chemistry of 2D intermetallic compounds. 
\end{abstract}


\maketitle

\section{Introduction}
Two-dimensional (2D) compound of CuAu has been synthesized experimentally \cite{zagler}, where 2D CuAu has a buckled honeycomb (BHC) structure, the same as the structure of silicene and germanene. This discovery opens the way for studying 2D compounds including only metallic elements. The fundamental properties of 2D elemental metals, such as energetic and structural properties \cite{nevalaita,hwang}, dynamical stability \cite{ono2020,ono2020_Po}, and magnetic properties \cite{ren}, have been investigated systematically from first-principles, also motivated by recent experimental synthesis of 2D metals \cite{wang2020}. It is desirable to extend such investigations for 2D intermetallic compounds that are difficult to be found (currently) in 2D materials databases \cite{ashton,choudhary,haastrup,feng,fukuda}.

Copper, silver, and gold belong to the noble metal column in the periodic table, so that the effect of $d$ electrons on their electronic and optical properties have been studied extensively under various conditions \cite{lin,bauer,xu,giri}. For the compounds including noble metals, it is not trivial to understand their stability properties among Cu$X$, Ag$X$, and Au$X$, where $X$ is an element in the periodic table, as confirmed by, for example, Pettifor's structure map \cite{pettifor} or modern materials database \cite{MP}. Recently, Alsalmi et al. have studied the stability of CuZn, AgZn, and AuZn in the B2 (CsCl-type) structure by using density functional theory (DFT) \cite{alsalmi} and explained nonmonotonic behaviors of structural parameters, where AgZn takes the largest lattice constant. This is never achieved by studying each compound individually. A systematic investigation for the stability of 2D Cu$X$, Ag$X$, and Au$X$ will thus be of fundamental importance. 

In a previous paper, the author studied a stability relationship between 2D and three-dimensional (3D) Cu-based compounds and found that if a Cu$X$ ($X$ is a metallic element) in the BHC structure is dynamically stable, that in the B$_h$ (WC-type) and L1$_1$ (CuPt-type) structures is also dynamically stable \cite{ono2021SR}. This is due to the geometric hierarchy: B$_h$ (L1$_1$) structure can be constructed from ABAB (ABC) stacking of hexagonal layers of Cu and $X$ alternately. In addition, it has been shown that an experimental synthesis of B$_h$ structure accounts for the dynamical stability of the BHC structure; for example, LiRh, LiPd, LiIr, and LiPt \cite{ono_satomi}. In this way, the BHC structure serves as a building block for constructing 3D compounds.  


In this paper, we show systematic investigations of the dynamical stability for BHC Cu$X$, Ag$X$, and Au$X$, where $X$ is a metallic element from Li to Pb in the periodic table, by using first-principles approach. We identify that more than 50 compounds are dynamically stable by quantifying the profile of the phonon spectra. We demonstrate that the quantity for the dynamical stability of Ag$X$ is correlated with that of Au$X$. Based on this fact, we predict that as a 3D compound AuZr in the B11 (CuTi-type) structure is dynamically stable. Trends of structural parameters, formation energy, electronic bands, and magnetic property are also discussed. We identify that the BHC AuK is a wide-gap semiconductors and that the BHC AgCr, AuCr, CuMn, AgMn, and AuMn have a magnetic moment about 4 $\mu_{\rm B}$ with a Bohr magneton $\mu_{\rm B}$. The calculated results are listed in Table \ref{table_ES} including the values of the modified formation energy defined in Eq.~(\ref{eq:form}) and the degree of the dynamical stability defined in Eq.~(\ref{eq:ratio}) and Table \ref{table_alat} including the lattice parameters of the BHC Cu$X$, Ag$X$, and Au$X$. 

\begin{table}
\begin{center}
\caption{The values of $E_{\rm form}$ in units of eV (the column 2, 3, and 4) and $S$ (the column 5, 6, and 7) for BHC $AX$. The value of $S$ is shown in bold if $S\ge 0.6$. }
{
\begin{tabular}{lrrrrrr} \hline\hline
$X$ \hspace{2mm} & \hspace{2mm} & $A$ \hspace{2mm} & \hspace{2mm} & \hspace{2mm} & $A$ \hspace{2mm} & \hspace{2mm} \\
\hspace{2mm} & Cu \hspace{2mm} & Ag \hspace{2mm} & Au \hspace{2mm} & Cu \hspace{2mm} & Ag \hspace{2mm} & Au \\
\hline
Li \hspace{2mm} & $-0.47$ \hspace{2mm} & $-0.64$ \hspace{2mm} & $-0.83$ \hspace{2mm} & ${\bf 1.00}$ \hspace{2mm} & ${\bf 0.82}$ \hspace{2mm} & ${\bf 0.98}$ \\ 
Na \hspace{2mm} & $0.40$ \hspace{2mm} & $-0.12$ \hspace{2mm} & $-0.19$ \hspace{2mm} & $-2.03$ \hspace{2mm} & $-0.53$ \hspace{2mm} & ${\bf 0.91}$ \\ 
K \hspace{2mm} & $0.71$ \hspace{2mm} & $0.01$ \hspace{2mm} & $-0.57$ \hspace{2mm} & $-2.22$ \hspace{2mm} & $-1.17$ \hspace{2mm} & ${\bf 1.00}$ \\ 
Rb \hspace{2mm} & $0.70$ \hspace{2mm} & $-0.00$ \hspace{2mm} & $-0.61$ \hspace{2mm} & $-2.75$ \hspace{2mm} & $-1.05$ \hspace{2mm} & $-1.11$ \\ 
Cs \hspace{2mm} & $0.65$ \hspace{2mm} & $-0.11$ \hspace{2mm} & $-0.71$ \hspace{2mm} & $-1.25$ \hspace{2mm} & $-2.54$ \hspace{2mm} & $-0.67$ \\ 
Be \hspace{2mm} & $-0.42$ \hspace{2mm} & $0.35$ \hspace{2mm} & $0.25$ \hspace{2mm} & ${\bf 0.97}$ \hspace{2mm} & $-3.86$ \hspace{2mm} & $-2.63$ \\ 
Mg \hspace{2mm} & $-0.24$ \hspace{2mm} & $-0.56$ \hspace{2mm} & $-0.66$ \hspace{2mm} & $-0.21$ \hspace{2mm} & ${\bf 0.89}$ \hspace{2mm} & ${\bf 0.97}$ \\ 
Ca \hspace{2mm} & $0.02$ \hspace{2mm} & $-0.40$ \hspace{2mm} & $-0.97$ \hspace{2mm} & $-3.01$ \hspace{2mm} & $-2.03$ \hspace{2mm} & $-1.51$ \\ 
Sr \hspace{2mm} & $0.19$ \hspace{2mm} & $-0.25$ \hspace{2mm} & $-0.90$ \hspace{2mm} & $-4.53$ \hspace{2mm} & $-3.28$ \hspace{2mm} & $-2.01$ \\ 
Ba \hspace{2mm} & $0.12$ \hspace{2mm} & $-0.28$ \hspace{2mm} & $-1.00$ \hspace{2mm} & $-3.57$ \hspace{2mm} & $-4.00$ \hspace{2mm} & $-1.64$ \\ 
Sc \hspace{2mm} & $-0.84$ \hspace{2mm} & $-1.01$ \hspace{2mm} & $-1.39$ \hspace{2mm} & $-2.19$ \hspace{2mm} & ${\bf 1.00}$ \hspace{2mm} & ${\bf 0.66}$ \\ 
Y \hspace{2mm} & $-0.70$ \hspace{2mm} & $-0.95$ \hspace{2mm} & $-1.47$ \hspace{2mm} & $-3.94$ \hspace{2mm} & $-1.19$ \hspace{2mm} & $-0.97$ \\ 
Lu \hspace{2mm} & $-0.89$ \hspace{2mm} & $-1.08$ \hspace{2mm} & $-1.66$ \hspace{2mm} & $-4.08$ \hspace{2mm} & $-0.86$ \hspace{2mm} & $-0.98$ \\ 
Ti \hspace{2mm} & $-1.07$ \hspace{2mm} & $-0.96$ \hspace{2mm} & $-1.13$ \hspace{2mm} & $0.01$ \hspace{2mm} & ${\bf 1.00}$ \hspace{2mm} & ${\bf 0.92}$ \\ 
Zr \hspace{2mm} & $-0.82$ \hspace{2mm} & $-0.97$ \hspace{2mm} & $-1.15$ \hspace{2mm} & $-3.05$ \hspace{2mm} & ${\bf 0.66}$ \hspace{2mm} & ${\bf 0.83}$ \\ 
Hf \hspace{2mm} & $-1.06$ \hspace{2mm} & $-1.11$ \hspace{2mm} & $-1.35$ \hspace{2mm} & $-3.13$ \hspace{2mm} & ${\bf 1.00}$ \hspace{2mm} & ${\bf 0.81}$ \\ 
V \hspace{2mm} & $-0.94$ \hspace{2mm} & $-0.43$ \hspace{2mm} & $-0.51$ \hspace{2mm} & $-4.29$ \hspace{2mm} & $-14.96$ \hspace{2mm} & $-24.01$ \\ 
Nb \hspace{2mm} & $-0.90$ \hspace{2mm} & $-0.79$ \hspace{2mm} & $-0.89$ \hspace{2mm} & $-1.60$ \hspace{2mm} & $-1.70$ \hspace{2mm} & $-1.22$ \\ 
Ta \hspace{2mm} & $-0.99$ \hspace{2mm} & $-0.84$ \hspace{2mm} & $-0.94$ \hspace{2mm} & $-3.19$ \hspace{2mm} & $-2.78$ \hspace{2mm} & $-1.76$ \\ 
Cr \hspace{2mm} & $-1.07$ \hspace{2mm} & $-0.67$ \hspace{2mm} & $-0.70$ \hspace{2mm} & $-3.47$ \hspace{2mm} & ${\bf 0.85}$ \hspace{2mm} & ${\bf 0.86}$ \\ 
Mo \hspace{2mm} & $-1.02$ \hspace{2mm} & $-0.64$ \hspace{2mm} & $-0.68$ \hspace{2mm} & $-2.97$ \hspace{2mm} & $-2.48$ \hspace{2mm} & $-2.18$ \\ 
W \hspace{2mm} & $-1.08$ \hspace{2mm} & $-0.70$ \hspace{2mm} & $-0.70$ \hspace{2mm} & $-2.31$ \hspace{2mm} & $-2.51$ \hspace{2mm} & $-3.08$ \\ 
Mn \hspace{2mm} & $-0.84$ \hspace{2mm} & $-0.72$ \hspace{2mm} & $-0.79$ \hspace{2mm} & ${\bf 1.00}$ \hspace{2mm} & ${\bf 1.00}$ \hspace{2mm} & ${\bf 0.76}$ \\ 
Tc \hspace{2mm} & $-1.07$ \hspace{2mm} & $-0.52$ \hspace{2mm} & $-0.47$ \hspace{2mm} & $-4.35$ \hspace{2mm} & $-5.01$ \hspace{2mm} & $-5.44$ \\ 
Re \hspace{2mm} & $-1.10$ \hspace{2mm} & $-0.52$ \hspace{2mm} & $-0.39$ \hspace{2mm} & $-2.26$ \hspace{2mm} & $-3.36$ \hspace{2mm} & $-3.72$ \\ 
Fe \hspace{2mm} & $-0.75$ \hspace{2mm} & $-0.22$ \hspace{2mm} & $-0.23$ \hspace{2mm} & $-2.63$ \hspace{2mm} & $-2.42$ \hspace{2mm} & $-3.12$ \\ 
Ru \hspace{2mm} & $-1.01$ \hspace{2mm} & $-0.50$ \hspace{2mm} & $-0.38$ \hspace{2mm} & $-1.28$ \hspace{2mm} & $0.12$ \hspace{2mm} & $0.55$ \\ 
Os \hspace{2mm} & $-0.99$ \hspace{2mm} & $-0.40$ \hspace{2mm} & $-0.15$ \hspace{2mm} & $-1.75$ \hspace{2mm} & $-1.36$ \hspace{2mm} & $-1.98$ \\ 
Co \hspace{2mm} & $-0.73$ \hspace{2mm} & $-0.03$ \hspace{2mm} & $0.03$ \hspace{2mm} & ${\bf 1.00}$ \hspace{2mm} & $-0.63$ \hspace{2mm} & $-0.29$ \\ 
Rh \hspace{2mm} & $-0.96$ \hspace{2mm} & $-0.60$ \hspace{2mm} & $-0.43$ \hspace{2mm} & ${\bf 0.75}$ \hspace{2mm} & ${\bf 0.83}$ \hspace{2mm} & ${\bf 0.85}$ \\ 
Ir \hspace{2mm} & $-0.90$ \hspace{2mm} & $-0.44$ \hspace{2mm} & $-0.13$ \hspace{2mm} & $-0.25$ \hspace{2mm} & $0.52$ \hspace{2mm} & $-0.77$ \\ 
Ni \hspace{2mm} & $-0.76$ \hspace{2mm} & $-0.12$ \hspace{2mm} & $-0.02$ \hspace{2mm} & ${\bf 0.91}$ \hspace{2mm} & $-0.17$ \hspace{2mm} & $-0.87$ \\ 
Pd \hspace{2mm} & $-0.89$ \hspace{2mm} & $-0.78$ \hspace{2mm} & $-0.56$ \hspace{2mm} & ${\bf 0.86}$ \hspace{2mm} & ${\bf 1.00}$ \hspace{2mm} & ${\bf 0.76}$ \\ 
Pt \hspace{2mm} & $-0.75$ \hspace{2mm} & $-0.54$ \hspace{2mm} & $-0.22$ \hspace{2mm} & ${\bf 0.79}$ \hspace{2mm} & ${\bf 0.83}$ \hspace{2mm} & ${\bf 0.97}$ \\ 
Cu \hspace{2mm} & $-0.51$ \hspace{2mm} & $-0.14$ \hspace{2mm} & $-0.09$ \hspace{2mm} & ${\bf 1.00}$ \hspace{2mm} & ${\bf 0.69}$ \hspace{2mm} & ${\bf 0.88}$ \\ 
Ag \hspace{2mm} & $-0.14$ \hspace{2mm} & $-0.33$ \hspace{2mm} & $-0.25$ \hspace{2mm} & ${\bf 0.69}$ \hspace{2mm} & ${\bf 0.75}$ \hspace{2mm} & ${\bf 0.88}$ \\ 
Au \hspace{2mm} & $-0.09$ \hspace{2mm} & $-0.25$ \hspace{2mm} & $-0.10$ \hspace{2mm} & ${\bf 0.88}$ \hspace{2mm} & ${\bf 0.88}$ \hspace{2mm} & ${\bf 0.87}$ \\ 
Zn \hspace{2mm} & $-0.41$ \hspace{2mm} & $-0.23$ \hspace{2mm} & $-0.25$ \hspace{2mm} & ${\bf 1.00}$ \hspace{2mm} & $0.50$ \hspace{2mm} & ${\bf 1.00}$ \\ 
Cd \hspace{2mm} & $0.02$ \hspace{2mm} & $-0.31$ \hspace{2mm} & $-0.30$ \hspace{2mm} & $-0.85$ \hspace{2mm} & ${\bf 0.82}$ \hspace{2mm} & ${\bf 0.93}$ \\ 
Hg \hspace{2mm} & $0.45$ \hspace{2mm} & $0.02$ \hspace{2mm} & $0.10$ \hspace{2mm} & $-3.29$ \hspace{2mm} & $-2.90$ \hspace{2mm} & $-2.18$ \\ 
Al \hspace{2mm} & $-0.72$ \hspace{2mm} & $-0.57$ \hspace{2mm} & $-0.64$ \hspace{2mm} & ${\bf 1.00}$ \hspace{2mm} & ${\bf 0.91}$ \hspace{2mm} & ${\bf 0.87}$ \\ 
Ga \hspace{2mm} & $-0.42$ \hspace{2mm} & $-0.44$ \hspace{2mm} & $-0.44$ \hspace{2mm} & ${\bf 0.65}$ \hspace{2mm} & ${\bf 0.62}$ \hspace{2mm} & ${\bf 0.83}$ \\ 
In \hspace{2mm} & $0.11$ \hspace{2mm} & $-0.29$ \hspace{2mm} & $-0.23$ \hspace{2mm} & $-3.24$ \hspace{2mm} & $-0.45$ \hspace{2mm} & $-0.85$ \\ 
Tl \hspace{2mm} & $0.42$ \hspace{2mm} & $-0.07$ \hspace{2mm} & $0.05$ \hspace{2mm} & $-5.67$ \hspace{2mm} & $-1.67$ \hspace{2mm} & $-3.17$ \\ 
Sn \hspace{2mm} & $0.07$ \hspace{2mm} & $-0.27$ \hspace{2mm} & $-0.19$ \hspace{2mm} & $-4.98$ \hspace{2mm} & $-0.62$ \hspace{2mm} & $-1.16$ \\ 
Pb \hspace{2mm} & $0.33$ \hspace{2mm} & $-0.11$ \hspace{2mm} & $0.04$ \hspace{2mm} & $-6.05$ \hspace{2mm} & $-2.44$ \hspace{2mm} & $-2.99$ \\ 
\hline\hline
\end{tabular}
}
\label{table_ES}
\end{center}
\end{table}

\begin{table}
\begin{center}
\caption{The values of $d(A\mathchar`-A)$ (the column 2, 3, and 4) and $d(A\mathchar`-X)$ (the column 5, 6, and 7) for BHC $AX$. The smaller value, $d(A\mathchar`-A)$ or $d(A\mathchar`-X)$, is underlined for each $AX$. }
{
\begin{tabular}{lrrrrrr} \hline\hline
$X$ \hspace{5mm} & \hspace{5mm} & $A$ \hspace{5mm} & \hspace{5mm} & \hspace{5mm} & $A$ \hspace{5mm} & \hspace{5mm} \\
\hspace{5mm} & Cu \hspace{5mm} & Ag \hspace{5mm} & Au \hspace{5mm} & Cu \hspace{5mm} & Ag \hspace{5mm} & Au \\
\hline
Li \hspace{5mm} &  2.63 \hspace{5mm} &   2.92 \hspace{5mm} &   2.91 \hspace{5mm} &   \underline{2.51} \hspace{5mm} &   \underline{2.68} \hspace{5mm} &  \underline{2.60} \\ 
Na \hspace{5mm} &  \underline{2.84} \hspace{5mm} &   \underline{3.06} \hspace{5mm} &   \underline{2.98} \hspace{5mm} &   2.93 \hspace{5mm} &   3.08 \hspace{5mm} &   2.99 \\ 
K  \hspace{5mm} &  5.55 \hspace{5mm} &   5.69 \hspace{5mm} &   5.56 \hspace{5mm} &   \underline{3.20} \hspace{5mm} &   \underline{3.29} \hspace{5mm} &   \underline{3.21} \\ 
Rb \hspace{5mm} &  5.75 \hspace{5mm} &   5.95 \hspace{5mm} &   5.78 \hspace{5mm} &   \underline{3.32} \hspace{5mm} &   \underline{3.44} \hspace{5mm} &   \underline{3.34} \\ 
Cs \hspace{5mm} &  5.94 \hspace{5mm} &   6.23 \hspace{5mm} &   5.98 \hspace{5mm} &   \underline{3.43} \hspace{5mm} &   \underline{3.60} \hspace{5mm} &   \underline{3.45} \\ 
Be \hspace{5mm} &  \underline{2.39} \hspace{5mm} &   2.69 \hspace{5mm} &   2.84 \hspace{5mm} &   2.40 \hspace{5mm} &   \underline{2.58} \hspace{5mm} &   \underline{2.38} \\ 
Mg \hspace{5mm} &  2.84 \hspace{5mm} &   2.99 \hspace{5mm} &   2.98 \hspace{5mm} &   \underline{2.72} \hspace{5mm} &   \underline{2.92} \hspace{5mm} &   \underline{2.83} \\ 
Ca \hspace{5mm} &  3.69 \hspace{5mm} &   3.72 \hspace{5mm} &   3.89 \hspace{5mm} &   \underline{2.87} \hspace{5mm} &   \underline{3.03} \hspace{5mm} &   \underline{2.88} \\ 
Sr \hspace{5mm} &  4.07 \hspace{5mm} &   4.13 \hspace{5mm} &   4.27 \hspace{5mm} &   \underline{3.03} \hspace{5mm} &   \underline{3.19} \hspace{5mm} &   \underline{3.04} \\ 
Ba \hspace{5mm} &  4.47 \hspace{5mm} &   4.49 \hspace{5mm} &   4.66 \hspace{5mm} &   \underline{3.14} \hspace{5mm} &   \underline{3.31} \hspace{5mm} &   \underline{3.17} \\ 
Sc \hspace{5mm} &  3.10 \hspace{5mm} &   3.14 \hspace{5mm} &   3.29 \hspace{5mm} &   \underline{2.67} \hspace{5mm} &   \underline{2.90} \hspace{5mm} &   \underline{2.74} \\ 
Y  \hspace{5mm} &  3.44 \hspace{5mm} &   3.43 \hspace{5mm} &   3.64 \hspace{5mm} &   \underline{2.81} \hspace{5mm} &   \underline{3.01} \hspace{5mm} &   \underline{2.86} \\ 
Lu \hspace{5mm} &  3.32 \hspace{5mm} &   3.33 \hspace{5mm} &   3.54 \hspace{5mm} &   \underline{2.74} \hspace{5mm} &   \underline{2.94} \hspace{5mm} &   \underline{2.78} \\ 
Ti \hspace{5mm} &  2.71 \hspace{5mm} &   \underline{2.85} \hspace{5mm} &   2.89 \hspace{5mm} &   \underline{2.66} \hspace{5mm} &   2.90 \hspace{5mm} &   \underline{2.80} \\ 
Zr \hspace{5mm} &  2.96 \hspace{5mm} &   \underline{3.00} \hspace{5mm} &   3.07 \hspace{5mm} &   \underline{2.77} \hspace{5mm} &   3.01 \hspace{5mm} &   \underline{2.89} \\ 
Hf \hspace{5mm} &  2.97 \hspace{5mm} &   3.01 \hspace{5mm} &   3.09 \hspace{5mm} &   \underline{2.72} \hspace{5mm} &   \underline{2.96} \hspace{5mm} &   \underline{2.83} \\ 
V  \hspace{5mm} &  \underline{2.51} \hspace{5mm} &   \underline{2.72} \hspace{5mm} &  \underline{2.75} \hspace{5mm} &   2.66 \hspace{5mm} &   2.87 \hspace{5mm} &   2.81 \\ 
Nb \hspace{5mm} &  \underline{2.71} \hspace{5mm} &   \underline{2.80} \hspace{5mm} &   \underline{2.82} \hspace{5mm} &   2.77 \hspace{5mm} &   3.00 \hspace{5mm} &   2.95 \\ 
Ta \hspace{5mm} &  2.75 \hspace{5mm} &   \underline{2.82} \hspace{5mm} &   \underline{2.85} \hspace{5mm} &   \underline{2.72} \hspace{5mm} &   2.97 \hspace{5mm} &   2.90 \\ 
Cr \hspace{5mm} &  \underline{2.44} \hspace{5mm} &   2.88 \hspace{5mm} &   2.87 \hspace{5mm} &   2.60 \hspace{5mm} &   \underline{2.86} \hspace{5mm} &   \underline{2.82} \\ 
Mo \hspace{5mm} &  \underline{2.63} \hspace{5mm} &   \underline{2.73} \hspace{5mm} &   \underline{2.75} \hspace{5mm} &   2.69 \hspace{5mm} &   2.94 \hspace{5mm} &   2.90 \\ 
W  \hspace{5mm} &  \underline{2.67} \hspace{5mm} &   \underline{2.74} \hspace{5mm} &   \underline{2.76} \hspace{5mm} &   2.67 \hspace{5mm} &   2.93 \hspace{5mm} &   2.90 \\ 
Mn \hspace{5mm} &  2.63 \hspace{5mm} &   \underline{2.83} \hspace{5mm} &   2.84 \hspace{5mm} &   \underline{2.59} \hspace{5mm} &   2.85 \hspace{5mm} &   \underline{2.82} \\ 
Tc \hspace{5mm} &  \underline{2.59} \hspace{5mm} &   \underline{2.69} \hspace{5mm} &   \underline{2.71} \hspace{5mm} &   2.64 \hspace{5mm} &   2.89 \hspace{5mm} &   2.87 \\ 
Re \hspace{5mm} &  \underline{2.61} \hspace{5mm} &   \underline{2.69} \hspace{5mm} &   \underline{2.71} \hspace{5mm} &   2.64 \hspace{5mm} &   2.90 \hspace{5mm} &   2.88 \\ 
Fe \hspace{5mm} &  \underline{2.48} \hspace{5mm} &   \underline{2.70} \hspace{5mm} &   \underline{2.73} \hspace{5mm} &   2.58 \hspace{5mm} &   2.83 \hspace{5mm} &   2.80 \\ 
Ru \hspace{5mm} &  \underline{2.58} \hspace{5mm} &   \underline{2.70} \hspace{5mm} &   \underline{2.72} \hspace{5mm} &   2.62 \hspace{5mm} &   2.88 \hspace{5mm} &   2.89 \\ 
Os \hspace{5mm} &  \underline{2.60} \hspace{5mm} &   \underline{2.68} \hspace{5mm} &   \underline{2.70} \hspace{5mm} &   2.63 \hspace{5mm} &   2.91 \hspace{5mm} &   2.91 \\ 
Co \hspace{5mm} &  \underline{2.44} \hspace{5mm} &   \underline{2.67} \hspace{5mm} &   \underline{2.70} \hspace{5mm} &   2.55 \hspace{5mm} &   2.77 \hspace{5mm} &   2.78 \\ 
Rh \hspace{5mm} &  \underline{2.60} \hspace{5mm} &   \underline{2.72} \hspace{5mm} &   \underline{2.73} \hspace{5mm} &   2.61 \hspace{5mm} &   2.86 \hspace{5mm} &   2.88 \\ 
Ir \hspace{5mm} &  \underline{2.61} \hspace{5mm} &   \underline{2.69} \hspace{5mm} &   \underline{2.71} \hspace{5mm} &   2.63 \hspace{5mm} &   2.91 \hspace{5mm} &   2.95 \\ 
Ni \hspace{5mm} &  \underline{2.45} \hspace{5mm} &   \underline{2.68} \hspace{5mm} &   \underline{2.71} \hspace{5mm} &   2.51 \hspace{5mm} &   2.72 \hspace{5mm} &   2.73 \\ 
Pd \hspace{5mm} &  2.65 \hspace{5mm} &   \underline{2.78} \hspace{5mm} &   \underline{2.77} \hspace{5mm} &   \underline{2.64} \hspace{5mm} &   2.84 \hspace{5mm} &   2.89 \\ 
Pt \hspace{5mm} &  2.66 \hspace{5mm} &   \underline{2.75} \hspace{5mm} &   \underline{2.74} \hspace{5mm} &   \underline{2.66} \hspace{5mm} &   2.89 \hspace{5mm} &   2.97 \\ 
Cu \hspace{5mm} &  \underline{2.50} \hspace{5mm} &   \underline{2.73} \hspace{5mm} &   \underline{2.74} \hspace{5mm} &   2.57 \hspace{5mm} &   2.76 \hspace{5mm} &   2.78 \\ 
Ag \hspace{5mm} &  \underline{2.73} \hspace{5mm} &   \underline{2.86} \hspace{5mm} &   \underline{2.82} \hspace{5mm} &   2.76 \hspace{5mm} &   2.98 \hspace{5mm} &   3.06 \\ 
Au \hspace{5mm} &  \underline{2.74} \hspace{5mm} &   \underline{2.82} \hspace{5mm} &   \underline{2.77} \hspace{5mm} &   2.78 \hspace{5mm} &   3.06 \hspace{5mm} &   3.30 \\ 
Zn \hspace{5mm} &  \underline{2.53} \hspace{5mm} &   \underline{2.76} \hspace{5mm} &   \underline{2.75} \hspace{5mm} &   2.74 \hspace{5mm} &   2.96 \hspace{5mm} &   2.99 \\ 
Cd \hspace{5mm} &  \underline{2.80} \hspace{5mm} &   \underline{2.91} \hspace{5mm} &   \underline{2.87} \hspace{5mm} &   2.91 \hspace{5mm} &   3.15 \hspace{5mm} &   3.18 \\ 
Hg \hspace{5mm} &  \underline{2.88} \hspace{5mm} &   \underline{2.99} \hspace{5mm} &   \underline{2.89} \hspace{5mm} &   2.93 \hspace{5mm} &   3.19 \hspace{5mm} &   3.34 \\ 
Al \hspace{5mm} &  2.65 \hspace{5mm} &   \underline{2.81} \hspace{5mm} &   2.83 \hspace{5mm} &   \underline{2.61} \hspace{5mm} &   2.86 \hspace{5mm} &   \underline{2.82} \\ 
Ga \hspace{5mm} &  \underline{2.65} \hspace{5mm} &   \underline{2.84} \hspace{5mm} &   \underline{2.83} \hspace{5mm} &   2.68 \hspace{5mm} &   2.96 \hspace{5mm} &   2.97 \\ 
In \hspace{5mm} &  2.98 \hspace{5mm} &   \underline{3.06} \hspace{5mm} &   \underline{3.01} \hspace{5mm} &   \underline{2.83} \hspace{5mm} &   3.09 \hspace{5mm} &   3.06 \\ 
Tl \hspace{5mm} &  3.26 \hspace{5mm} &   3.19 \hspace{5mm} &   3.36 \hspace{5mm} &   \underline{2.84} \hspace{5mm} &   \underline{3.17} \hspace{5mm} &   \underline{2.98} \\ 
Sn \hspace{5mm} &  3.13 \hspace{5mm} &   3.13 \hspace{5mm} &   3.12 \hspace{5mm} &   \underline{2.76} \hspace{5mm} &   \underline{3.05} \hspace{5mm} &   \underline{2.98} \\ 
Pb \hspace{5mm} &  3.33 \hspace{5mm} &   3.26 \hspace{5mm} &   3.24 \hspace{5mm} &   \underline{2.80} \hspace{5mm} &   \underline{3.12} \hspace{5mm} &   \underline{3.05} \\ 
\hline\hline
\end{tabular}
}
\label{table_alat}
\end{center}
\end{table}

\section{Basic concepts}
\begin{figure}
\center
\includegraphics[scale=0.4]{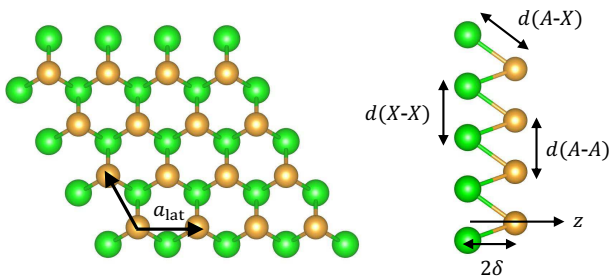}
\caption{Top and side views for the 2D $AX$ in the BHC structure with the lattice constant $a_{\rm lat}$, the thickness $2\delta$, and the interatomic distances $d(A\mathchar`-A)$, $d(A\mathchar`-X)$, and $d(X\mathchar`-X)$. } \label{fig_1} 
\end{figure}

The primitive lattice vectors of the BHC structure are $\bm{a}_1=a_{\rm lat}(1,0,0)$, $\bm{a}_2=a_{\rm lat}(-1/2,\sqrt{3}/2,0)$, and $\bm{a}_3=(0,0,c)$ with the lattice constant $a_{\rm lat}$. For the BHC $AX$ with $A=$ Cu, Ag, and Au and $X$ a metallic element, the basis vectors are $A(0,0,\delta)$ and $X(0,a_{\rm lat}/\sqrt{3},-\delta)$ with the total thickness $2\delta$. The interatomic distance $d(A\mathchar`-A)$ between the same species is equal to $a_{\rm lat}$, while the interatomic distance $d(A\mathchar`-X)$ between atoms $A$ and $X$ is given by $a_{\rm lat}\sqrt{1/3+4r^2}$ with $r=\delta/a_{\rm lat}$ (see Fig.~\ref{fig_1}). 

The formation energy is usually defined as the energy difference between the total energy of the binary system considered (such as ordered compounds and/or alloys) and that of the reference system (ground state structure of elements, such as fcc, bcc, and hcp). Application of this standard formula to 2D binary compounds yielded positive formation energies for most cases because 2D structures are energetically unstable (but might be dynamically stable) \cite{ono_satomi}. To obtain a useful insight from the energy calculations, we define a modified formation energy as
\begin{eqnarray}
 E_{\rm form}(AX) =  \varepsilon_{\rm BHC}(AX) - \left[  \varepsilon_{\rm HX}(A)  +  \varepsilon_{\rm HX}(X)  \right],
 \label{eq:form}
\end{eqnarray}
where $\varepsilon_{\rm BHC}(AX)$ is the total energy of $AX$ in the BHC structure and $\varepsilon_{\rm HX}(A)$ is the total energy of $A$ in the hexagonal (HX) structure. The physics behind Eq.~(\ref{eq:form}) is that vertically stacking the HX layers of $A$ and $X$ produces the BHC structure of $AX$ by releasing or absorbing energies: When $E_{\rm form}(AX)<0$, BHC $AX$ is energetically suitable to be formed from HX layers. 

For the elemental metals of $X$, 46 elements were investigated in the present work, so that the phonon dispersions for 135 compounds were calculated. To systematically understand the dynamical stability property of $AX$, we define the degree of dynamical stability $S$ as
\begin{eqnarray}
  S = \frac{R - (1-p)}{p}
 \label{eq:ratio}
\end{eqnarray}
with 
\begin{eqnarray}
 R= \frac{\omega_{\rm max} + {\rm sgn} (\omega_{\rm min}^{2}) \vert \omega_{\rm min}\vert }{\omega_{\rm max}},
\end{eqnarray}
where $\omega_{\rm max}$ and $\omega_{\rm min}$ are the maximum and minimum phonon frequencies, respectively, which are calculated by phonon density-of-states (DOS). An imaginary frequency is represented as a negative frequency by using the sign function of ${\rm sgn}$. If no imaginary frequency is observed in the phonon DOS, $R=1$ exactly, otherwise $R<1$. Due to a finite size of $q$ grid in the Brillouin zone, small negative frequencies may appear around the $\Gamma$ point. In order to identify an unstable structure clearly, we subtract $1-p$ from $R$ and normalize the value by dividing $p$. In the present work, we adapted the value of $p=0.1$. The negative (or small positive) value of $S$ indicates an instability of $AX$ in the BHC structure.  


\section{Results and Discussion}

\begin{figure}
\center
\includegraphics[scale=0.45]{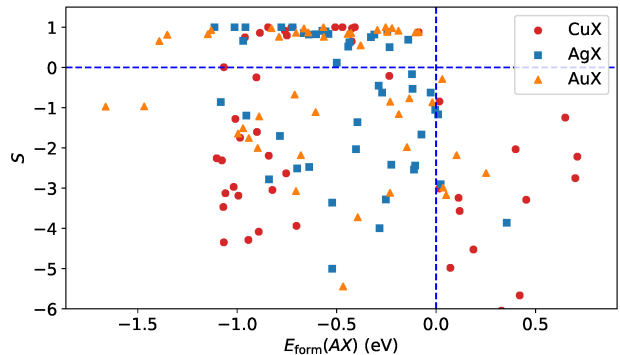}
\caption{Plot of $S$ as a function of $E_{\rm form}$ for the BHC Cu$X$, Ag$X$, and Au$X$. Dashed lines indicate $E_{\rm form}=0$ (vertical) and $S=0$ (horizontal).} \label{fig_2} 
\end{figure}

\begin{figure*}
\center
\includegraphics[scale=0.5]{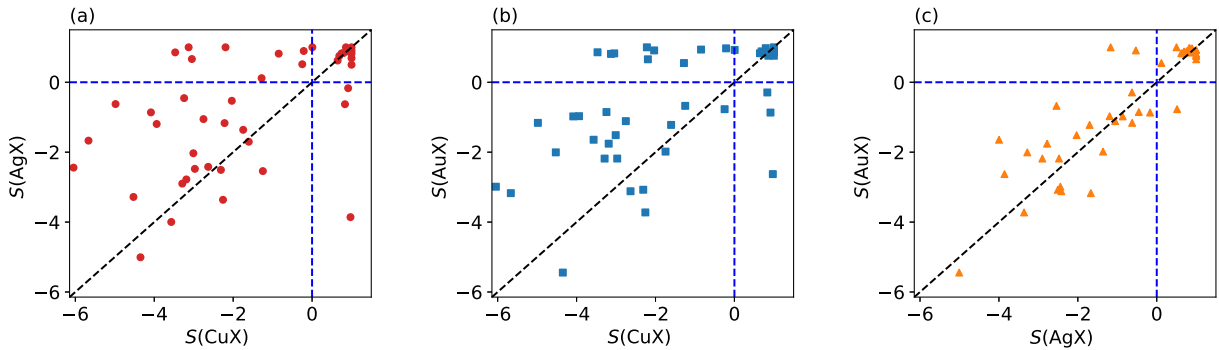}
\caption{Relationship of $S$ between $AX$ and $BX$: (a) $(A,B)=$ (Cu, Ag), (b) $(A,B)=$ (Cu, Au), and (c) $(A,B)=$ (Ag, Au). Diagonal line (dashed) indicates $S(AX)=S(BX)$. } \label{fig_3} 
\end{figure*}

\begin{figure}
\center
\includegraphics[scale=0.45]{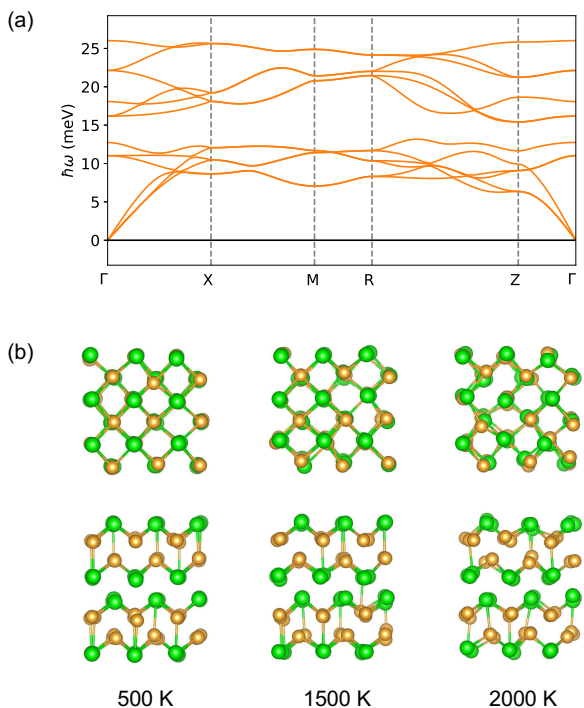}
\caption{(a) The phonon dispersions of AuZr in the B11 structure and (b) the atomic distribution of the 3$\times$2$\times$2 supercell after the first-principles MD simulation of 3.2 ps under 500, 1500, and 2000 K. The views are from the $c$ axis (top) and $a$ axis (bottom). } \label{fig_4} 
\end{figure}

\subsection{Dynamical stability}
\label{sec:DS}
Table \ref{table_ES} lists the values of $E_{\rm form}(AX)$ in Eq.~(\ref{eq:form}) and $S$ in Eq.~(\ref{eq:ratio}) for Cu$X$, Ag$X$, and Au$X$ in the BHC structure. The values of $S$ are distributed from $-6$ (unstable) to $1$ (stable), except for AgV and AuV. The BHC $AX$ having relatively high value of $S$ will be a cousin of 2D CuAu \cite{zagler}, which will expand the family of 2D intermetallic compounds. The calculated phonon dispersions for Cu$X$, Ag$X$, and Au$X$ are provided in the Supplemental Material. 

We first study a correlation between $E_{\rm form}(AX)$ and $S$. As shown in Fig.~\ref{fig_2}, the $AX$ having positive $S$ shows negative $E_{\rm form}(AX)$, indicating that the formation energy modified to treat an energy gain for stacking HX layers, defined as Eq.~(\ref{eq:form}), may be a good quantity for studying the dynamical stability of $AX$. This improves our previous results \cite{ono_satomi}, where the 2D compound that is dynamically stable can have a positive formation energy (even in 2D CuAu) when the standard formation energy is applied. However, negative $E_{\rm form}(AX)$ is not still a sufficient condition for positive $S$, which is left for an open question.

As shown in Fig.~\ref{fig_2}, the distribution of $E_{\rm form}({\rm Cu}X)$ seems to be different from that of $E_{\rm form}({\rm Ag}X)$ and $E_{\rm form}({\rm Au}X)$. To understand the dynamical stability relationship among 2D compounds, we plot $S$ for Ag$X$ and Au$X$ as a function of Cu$X$ in Fig.~\ref{fig_3}(a) and \ref{fig_3}(b), respectively. The trends of the dynamical stability for Cu$X$ is not similar to those for Ag$X$ and Au$X$ because the plots of $S$ are scattered around dashed lines $S({\rm Cu}X)=S({\rm Ag}X)$ and $S({\rm Cu}X)=S({\rm Au}X)$. In addition, even if Cu$X$ in the BHC structure is dynamically stable, i.e., $S({\rm Cu}X)\simeq 1$, Ag$X$ and Au$X$ in the BHC structure are unstable, i.e., $S({\rm Ag}X)\ll 1$ or $S({\rm Au}X)\ll 1$, and vice versa: For $X=$ Be, Co, and Ni, Cu$X$ is dynamically stable, while Ag$X$ and Au$X$ are unstable. On the other hand, for $X=$ Mg, Sc, Ti, Zr, Hf, Cr, and Cd, Ag$X$ and Au$X$ are dynamically stable, while Cu$X$ is unstable. These anomalies can be explained by negatively large values of $E_{\rm form}(AX)$ for $X=$ Be, Co, Ni, Mg, Sc, and Cd (see Table \ref{table_ES}). An interesting finding is that the value of $S$ for Ag$X$ is well correlated with that of Au$X$, as shown in Fig.~\ref{fig_3}(c). The correlation coefficient between Ag$X$ and Au$X$ is 0.94, whereas that between Cu$X$ and Ag$X$ (Au$X$) is 0.49 (0.42). The strong correlation of $S$ between Ag$X$ and Au$X$ is attributed to the similar distribution of $E_{\rm form}$ in Fig.~\ref{fig_2}. 


Combined with the correlation of $S$ between Ag$X$ and Au$X$ and the stability relationship between 2D and 3D structures, we can predict stable compounds that have not been synthesized yet experimentally. Let us consider the cases of $X=$ Ti, Zr, and Hf, where Ag$X$ and Au$X$ are dynamically stable as listed in Table \ref{table_ES}. By using Atom Works database \cite{nims}, we found that AgTi \cite{AgTi}, AgZr \cite{AgZr}, AgHf \cite{AgHf}, AuTi \cite{AuTiAuHf}, and AuHf \cite{AuTiAuHf} in the B11 structure have already been synthesized experimentally. We thus identify an anomaly that no synthesis of AuZr in the B11 structure has been reported so far. We then calculated the phonon dispersions of AuZr, as shown in Fig.~\ref{fig_4}(a). No imaginary phonon frequencies are observed along the symmetry lines. To study the temperature effect on the dynamical stability, we also performed first-principles molecular dynamics (MD) simulation. The B11 AuZr is stable up to 1500 K (see Fig.~\ref{fig_4}(b)). 

It is quite interesting that the synthesis probability of B11 AuZr has been predicted to be 91.1 \%, according to the MaterialNet program \cite{aykol}. This synthesizability was derived from machine learning the time evolution of materials stability network, where the network is constructed from the convex free-energy surface of inorganic materials. In this way, different approaches have predicted the stability of B11 AuZr, so that we expect that AuZr in the B11 structure will be potentially synthesized in future experiments. 

By considering the stability relationship between 2D and 3D structures again, the positive values of $S$ for BHC AgPd, AgPt, and CuPt listed in Table \ref{table_ES} are consistent with the prediction by Nelson et al. \cite{nelson}, where AgPd, AgPt, and CuPt have the L1$_1$ structure as its ground state. This is because if the BHC structure is dynamically stable, then the L1$_1$ structure is also stable \cite{ono2021SR}. An implication in this section is that the 2D and 3D structures must be interrelated {\it via the dynamical stability}. A similar interrelation can hold for the Ag-based and Au-based compounds. 

\subsection{Structure relationship}
\label{sec:structure}
We next study how the structural parameters of the BHC structure are related to those of the 3D counterparts. To investigate it, we list the values of $d(A\mathchar`-A)$ and $d(A\mathchar`-X)$ in Table \ref{table_alat}. If the value of $d(A\mathchar`-A)$ is smaller than that of $d(A\mathchar`-X)$, the coordination number is equal to six because of a strong intralayer bonding ($A$-$A$ and $X$-$X$). In contrast, if an interlayer bonding ($A$-$X$) is rather strong, an inequality $d(A\mathchar`-A)/d(A\mathchar`-X)>1$ will hold, leading to smaller coordination number of three. 

We consider the cases of $X=$ Cu, Ag, and Au, where an inequality $d(A\mathchar`-A)/d(A\mathchar`-X)<1$ holds. Let us just remind that the noble metals (Cu, Ag, and Au) prefer to have the fcc structure with the largest coordination number of twelve. Therefore, the inequality of $d(A\mathchar`-A)/d(A\mathchar`-X)<1$ for the 2D Cu, Ag, and Au in the BHC structure reflects the close-packed structure in the 3D Cu, Ag, and Au. By using Atom Works database \cite{nims}, we found that the mixing of noble metals produces a solid solution in the fcc structure represented by Cu$_{0.5}$Ag$_{0.5}$ \cite{linde}, Cu$_{0.5}$Au$_{0.5}$ \cite{nahm}, and Ag$_{0.5}$Au$_{0.5}$ \cite{venudhar}. These support the high coordination number in 2D CuAg, CuAu, and AgAu. 

\begin{figure}
\center
\includegraphics[scale=0.45]{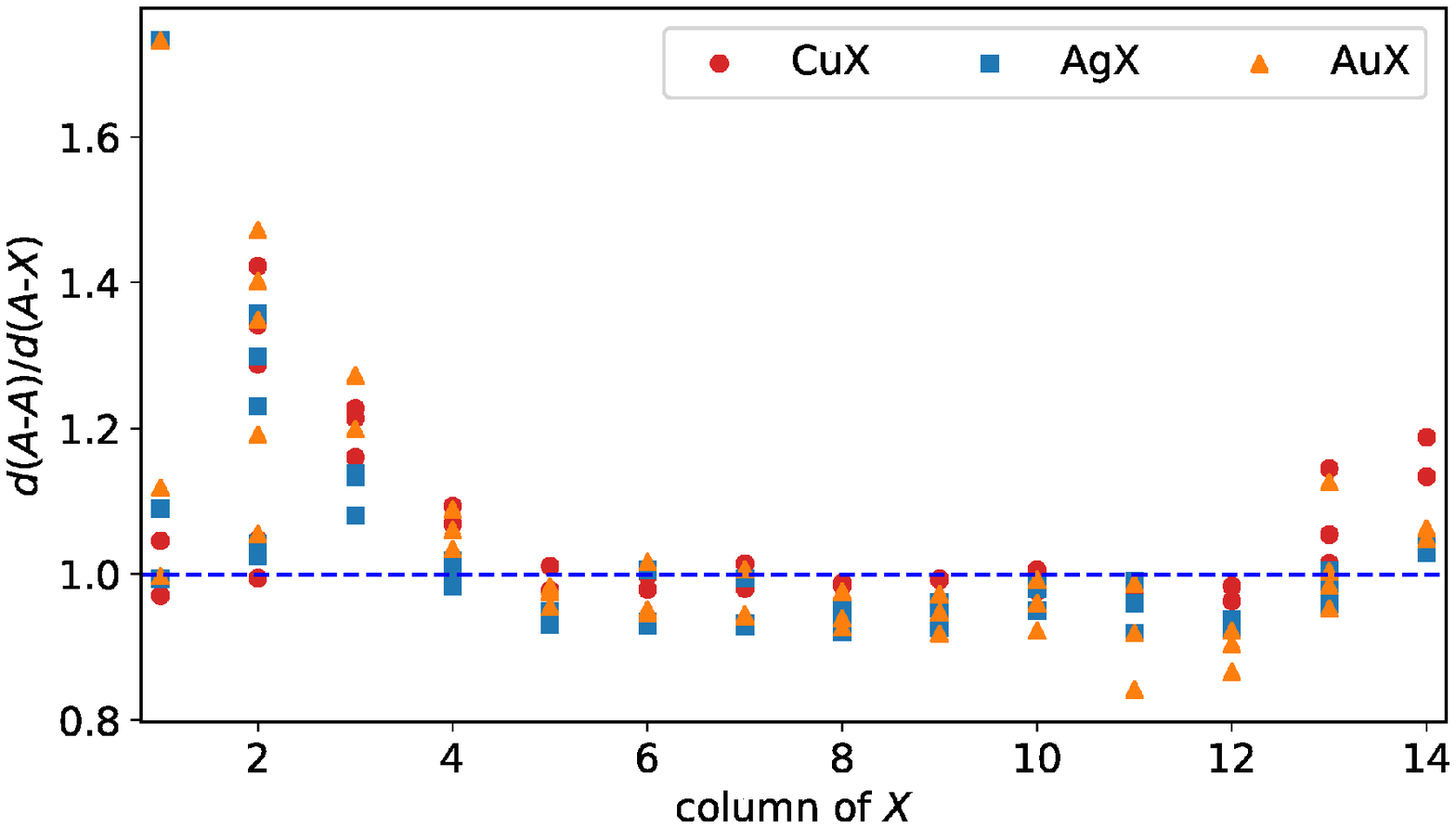}
\caption{Plot of $d(A\mathchar`-A)/d(A\mathchar`-X)$ versus the column of $X$ in the periodic table for the BHC Cu$X$, Ag$X$, and Au$X$.} \label{fig_5} 
\end{figure}

To study the structural trends for the $AX$ in the BHC structure, we plot the values of $d(A\mathchar`-A)/d(A\mathchar`-X)$ versus the column of $X$ in the periodic table, as shown in Fig.~\ref{fig_5}. When $X$ is an element of the column 1, 2, 3, and 4, an inequality $d(A\mathchar`-A)>d(A\mathchar`-X)$ holds. (The $AX$s including alkali metals of K, Rb and Cs have $d(A\mathchar`-A)/d(A\mathchar`-X)=\sqrt{3}$, i.e., no buckling structure.) Again, let us just remind that noble metals can mix with some elements of the column 1, 2, and 3 to form the B2 structure, such as AuRb and AuCs \cite{tinelli}, AgMg \cite{AgMg}, AuMg \cite{AuMgAuCd}, and AgSc and AuSc \cite{Sc}, where the coordination number is eight that is smaller than twelve. We can attribute the inequality of $d(A\mathchar`-A)/d(A\mathchar`-X)>1$ partially to the formation of the B2 structure of $AX$. When the column of $X$ is from 5 to 10, the magnitude of $d(A\mathchar`-A)$ is smaller than that of $d(A\mathchar`-X)$, implying a strong intralayer bonding as in the BHC Cu, Ag, and Au (the column 11). However, for $X=$ the column 12, it is difficult to rationalize the inequality $d(A\mathchar`-A)<d(A\mathchar`-X)$ because $A$Zn, AgCd \cite{AgCd}, and AuCd \cite{AuMgAuCd} in the B2 structure have been synthesized, as for $X=$ the column of 1, 2, and 3. For $X=$ the column 13 and 14 elements, no compounds of the forms of $AX$ and $A_{\rm 0.5}X_{\rm 0.5}$ have been reported except for $X=$ Al and Ga. 

It will be useful to study trends of structural parameters with $X$ fixed. Recently, Alsalmi et al. have studied the stability of CuZn, AgZn, and AuZn in the B2 structure by using DFT \cite{alsalmi}. They have shown that the bonding strength between noble metal and Zn atom becomes stronger as the $d$ orbitals are more extended. In contrast, the core-core repulsion enhances the lattice constant, in turn, giving rise to a decrease in the overlap of wavefunctions. These effects yield nonmonotonic stability properties as one goes CuZn, AgZn, and AuZn and a large lattice parameter of AgZn. In the present calculations, similar behaviors can be observed in BHC $AX$: an inequality $d({\rm Cu}\mathchar`-{\rm Cu})<d({\rm Au}\mathchar`-{\rm Au})<d({\rm Ag}\mathchar`-{\rm Ag})$ or $d({\rm Cu}\mathchar`-X)<d({\rm Au}\mathchar`-X)<d({\rm Ag}\mathchar`-X)$ can hold. For the cases of $X=$ Tl, Sn, and Pb, the relationship between $d({\rm Cu}\mathchar`-{\rm Cu})$, $d({\rm Ag}\mathchar`-{\rm Ag})$, and $d({\rm Au}\mathchar`-{\rm Au})$ is quite different. 


\subsection{Electronic and magnetic properties}
Toward practical applications, bandgap engineering of 2D materials has been studied by introducing an electric field, strain, and foreign atoms \cite{chaves}. It is noteworthy that for 3D compounds, AuRb and AuCs in the B2 structure have been known as semiconductors \cite{spicer}, although both Rb and Cs are metallic elements. Recently, Adeleke et al. have predicted that at high pressure K$_2$Ni is a semiconductor with an indirect bandgap of 0.65 eV, which is due to a Peierls gap opening \cite{adeleke}. It will be interesting to study which 2D compounds including two metallic elements may exhibit a semiconducting property. 

The spin-dependent electron band structure for Cu$X$, Ag$X$, and Au$X$ in the BHC structure is provided in the Supplemental Material. Basically, nearly flat $d$-bands are located below the Fermi level by a few eV. As in the band structure of elemental metals, the location of $d$-bands tend to be shallow and deep for Cu$X$ and Ag$X$, respectively. The $s$ electron band is modified by the mixing with other electrons of $X$ as well as the $s$-$d$ coupling and crosses the Fermi level. 

\begin{figure}
\center
\includegraphics[scale=0.43]{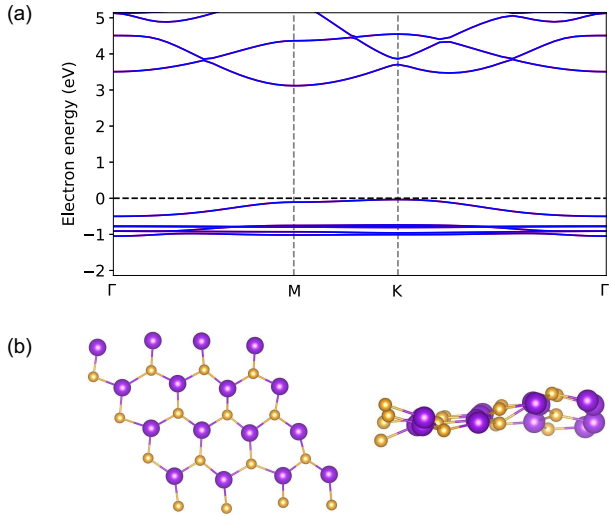}
\caption{(a) The electron band structure of AuK in the BHC structure and (b) the top and side views of the atomic distribution of the 4$\times$4$\times$1 supercell (32 atoms) after the first-principles MD simulation of 1 ps under 150 K, indicating an instability against out-of-plane vibrations. } \label{fig_6} 
\end{figure}

We identify that the compounds with alkali metals $X=$ K, Rb, and Au are semiconductors. Figure \ref{fig_6}(a) show the electron band structure of AuK, showing an indirect semiconductor with a bandgap of 3.16 eV, where the valence band maximum and conduction band minimum are located at K and M point, respectively. It is well known that the DFT within the generalized gradient approximation underestimates the magnitude of the bandgap. In this way, BHC AuK can be included to one of the wide-gap semiconductors such as hexagonal BN \cite{2DBN} and 2D GaN \cite{2DGaN}. Although AuK in the BHC structure is predicted to be dynamically stable at zero temperature, first-principles MD simulation predicts that it is unstable even at low temperature of 150 K, as shown in Fig.~\ref{fig_6}(b). While the hexagonal symmetry is almost preserved (top view), the out-of-plane displacement is significantly large (side view) at 1 ps, giving rise to a failure of self-consistent field convergence. The instability against the temperature is due to small phonon energies (a few meV) of the flexural modes around the M and K points (see Fig.~S1 in the Supplemental Material). It is highly desirable to propose a substrate material for stabilizing the BHC AuK. 

\begin{table}[bb]
\begin{center}
\caption{The magnetic moment (in units of $\mu_{\rm B}$ per a cell) of Cu$X$, Ag$X$, and Au$X$ in the BHC structure. The magnetic moments for the dynamically stable structures ($S\ge 0.6$) are shown in bold. }
{
\begin{tabular}{lrrr} \hline\hline
$X$ \hspace{7mm} & \hspace{7mm} & $A$ \hspace{7mm} &  \\ 
  \hspace{7mm} & Cu \hspace{7mm} & Ag \hspace{7mm} & Au \\
\hline
Sr \hspace{7mm} & 0.17 \hspace{7mm} & 0.14 \hspace{7mm} & 0.34  \\
Ba \hspace{7mm} & 0.66 \hspace{7mm} & 0.61 \hspace{7mm} & 0.65  \\
Ti \hspace{7mm} & 0.87 \hspace{7mm} & {\bf 1.01} \hspace{7mm} & {\bf 1.14}  \\
Zr \hspace{7mm} & 0.70 \hspace{7mm} & {\bf 0.64} \hspace{7mm} & {\bf 0.82}  \\
Hf \hspace{7mm} & 0.75 \hspace{7mm} & {\bf 0.75} \hspace{7mm} & {\bf 0.90}  \\
V \hspace{7mm} & 0.02 \hspace{7mm} & 1.13 \hspace{7mm} & 0.88  \\
Cr \hspace{7mm} & 0.00 \hspace{7mm} & {\bf 4.29} \hspace{7mm} & {\bf 4.28}  \\
Mn \hspace{7mm} & {\bf 3.85} \hspace{7mm} & {\bf 4.20} \hspace{7mm} & {\bf 4.31}  \\
Fe \hspace{7mm} & 2.64 \hspace{7mm} & 3.00 \hspace{7mm} & 3.09  \\
Ru \hspace{7mm} & 0.01 \hspace{7mm} & 1.19 \hspace{7mm} & 1.42  \\
Os \hspace{7mm} & 0.00 \hspace{7mm} & 0.01 \hspace{7mm} & 0.38  \\
Co \hspace{7mm} & {\bf 1.74} \hspace{7mm} & 1.86 \hspace{7mm} & 1.95 \\
Rh \hspace{7mm} & {\bf 0.59} \hspace{7mm} & {\bf 0.82} \hspace{7mm} & {\bf 0.98}  \\
Ir \hspace{7mm} & 0.01 \hspace{7mm} & 0.17 \hspace{7mm} & 0.71  \\
Ni \hspace{7mm} & {\bf 0.46} \hspace{7mm} & 0.61 \hspace{7mm} & 0.72  \\
\hline\hline
\end{tabular}
}
\label{table3}
\end{center}
\end{table}


We move on to the study of magnetic property. In general, reduced dimensionality can have a possibility for enhancing the magnitude of magnetic moments \cite{vaz}. It has been recently predicted that 18 elemental metals show a finite magnetic moment in the 2D structures, while there are only 5 magnetic elements in the 3D structures \cite{ren}. Also, the Cu-based compounds in the BHC structure exhibit a relatively high magnetic moment, compared to the 3D counterparts \cite{ono2021SR}. The replacement of Cu with Ag or Au in the 2D compounds will change a bonding character between adjacent atoms and the degree of $d$ electron localization as well, leading to an increase and/or decrease in the magnetic moment. 

Table \ref{table3} lists the magnetic moment of selected compounds $AX$ in the BHC structure. Even when nonmagnetic elements ($X=$ Sr, Ba, Ti, Zr, Hf, V, Ru, Os, Rh, and Ir) are included, finite magnetic moments appears due to the different profile of electron bands between up and down spins (see the Supplemental Material). This tendency is similar to the DFT calculations for 2D elemental metals \cite{ren}. The Cu$X$ tends to have smaller magnetic moment than Ag$X$ and Au$X$. The following interpretation will be applied: Among Cu, Ag, and Au, the lattice parameters of Cu$X$ is smallest, as listed in Table \ref{table_alat} and discussed in Sec.~\ref{sec:structure}; the hybridization of the orbital between different atoms will be strong in Cu$X$; and the delocalization of $d$ electron yields a decrease in the band splitting. 

The coupling between the magnetism and the dynamical stability has long been investigated for elemental metals \cite{grimvall}. An interesting feature in the present calculations is that the 2D compounds including anti-ferromagnetic elements of Cr and Mn have large magnetic moment about 4 $\mu_{\rm B}$ and are also dynamically stable. On the other hand, the 2D compounds with ferromagnetic elements of Fe, Co, and Ni have finite magnetic moments but are unstable except for CuCo and CuNi. The latter is similar to the tendency observed in our previous calculations \cite{ono2020}, where magnetic effects lead to an instability of 2D Fe and Co in the buckled square structure.

In our previous work \cite{ono_satomi}, we predicted that the BHC AuTi is unstable against the excitation of phonons at M and K points, by performing spin-unpolarized calculations. In the present work, BHC AuTi have finite magnetic moment listed in Table \ref{table3} and is dynamically stable. The magnetic effect on the dynamical stability of $AX$ seems to strongly depend on the column of $X$. 

\section{Summary}
We systematically investigated the dynamical stability and electronic property of BHC Cu$X$, Ag$X$, and Au$X$, where $X$ is 46 metallic element in the periodic table, by using first-principles approach. More than 50 compounds were identified to be dynamically stable by quantifying the profile of the phonon spectra. We demonstrated that the dynamical stability property of Ag$X$ is correlated with that of Au$X$. We predicted that as a 3D compound AuZr in the B11 structure is dynamically stable. This was speculated by a strong correlation of the quantity for the dynamical stability between Ag$X$ and Au$X$. Future work is to explore another interrelationship of structural and dynamical properties between the 2D and 3D compounds through more comprehensive search in the periodic table and/or other structures (including solid solutions) and with the use of machine learning approach. 

In addition, we calculated the electronic and magnetic properties of BHC $AX$. The AuK was predicted to be a wide-gap semiconductor, while a substrate material is needed to stabilize the out-of-plane vibrations. The AgCr, AuCr, CuMn, AgMn, and AuMn were predicted to show a ferromagnetic phase with a large magnetic moment. An interesting but complex relationship between the magnetism and the dynamical stability was also identified. 

\appendix
\section{Methods}
We optimized the parameters $a_{\rm lat}$ and $\delta$ for BHC $AX$ by using the DFT implemented to Quantum ESPRESSO (QE) package \cite{qe}. We used ultrasoft pseudo-potential \cite{dalcorso} and exchange-correlation functional parametrized by Perdew, Burke, and Ernzerhof within the generalized gradient approximation \cite{pbe}. The self-consistent field (scf) calculations were performed by setting the energy cutoff to be 80 Ry and 800 Ry for wavefunction and charge density, respectively, and the energy convergence to be $10^{-10}$ Ry. The $k$ grid was assumed to be 30$\times$30$\times$1 \cite{MK} with the smearing parameter of 0.02 Ry \cite{smearing}. The value of $c$ in $\bm{a}_3$ was set to be 14 \AA \ to avoid spurious interactions between different cells along the $z$ axis. The threshold values for geometry optimization were set to be $10^{-5}$ Ry for the total energy and $10^{-4}$ a.u. for the total force. Spin-polarized calculations were performed for all compounds, where a ferromagnetic phase was assumed as an initial guess of the scf calculations. The parameters (the size of the $k$ grid and the threshold values for geometry optimization) used in the present work are equal to those used in the previous work for elemental metals \cite{ono2020}.

We next calculated the phonon dispersions along the symmetry lines of $\Gamma$-M-K-$\Gamma$ by using density-functional perturbation theory (DFPT) \cite{dfpt} implemented to QE. The parameters for the scf calculations were the same as the geometry optimization calculations above. The convergence parameter for the DFPT calculations was set to be tr2\_ph=10$^{-14}$. The $q$ grid was assumed to be 6$\times$6$\times$1 (seven $q$ points in the irreducible Brillouin zone). By diagonalizing the dynamical matrix, we finally obtain the phonon frequencies $\omega_\alpha(\bm{q})$ at the wavevector $\bm{q}$ in the $\alpha$th band with $\alpha=1,\cdots,6$. If $\omega_\alpha(\bm{q})$ is imaginary, the BHC structure is unstable against such a phonon mode. 

By using the QE\cite{qe}, we optimized the geometry of B11 AuZr with the use of the cutoff energies of 60 and 600 Ry for wavefunction and charge density, respectively, and 16$\times$16$\times$16 $k$ grid. The optimized lattice parameters were $a=3.554$ \AA \ and $c=6.391$ \AA \ for the tetragonal unit cell. The phonon dispersion calculation was done by assuming 3$\times$3$\times$3 $q$ grid. We also performed first-principles MD simulation implemented in QE \cite{qe}. We used a 3$\times$3$\times$2 supercell (72 atoms) by considering the $\Gamma$ point in the Brillouin zone and set the ionic temperature of 500, 1000, 1500, and 2000 K by using the velocity scaling method. The time step of 20 a.u. (0.96 fs) was adapted and 3300 MD steps were done, corresponding to 3.2 ps.  

\begin{acknowledgments}
This study was supported by the a Grant-in-Aid for Scientific Research (C) (Grant No. 21K04628) from JSPS. The computation was carried out using the facilities of the Supercomputer Center, the Institute for Solid State Physics, the University of Tokyo.
\end{acknowledgments}





\end{document}